\documentclass[aip,pop,12pt,reprint]{revtex4-1}
\usepackage[dvips]{graphicx}
\usepackage{here}
\usepackage{pgfgantt}
 \usepackage{delarray}
\usepackage{amsmath}
\usepackage{amssymb}
\newcommand{\g}[1]{\mbox{\boldmath $#1$}}

\usepackage{ae}
\usepackage{color}
\usepackage{ifthen}
\usepackage{bm}
\usepackage{ifpdf}
\ifpdf
	\usepackage[pdftex]{hyperref}
	\hypersetup{colorlinks=true,
		pdfstartview=FitV,
		linkcolor=blue,
		citecolor=red,
		urlcolor=blue,
	}
	\DeclareGraphicsExtensions{.pdf}
\else
	\usepackage[dvips]{graphicx}
	\usepackage{epsfig}
	\DeclareGraphicsExtensions{.eps}
	\usepackage{url}
\fi

\newcommand{\beq}{\begin{equation}}
\newcommand{\eeq}{\end{equation}}

\usepackage{units}

\begin{document}

\title{Prospects and limitations of wakefield acceleration in solids}

\author{B. Svedung Wettervik}
\affiliation{Department of Physics, Chalmers University of Technology, G\"{o}teborg, Sweden}
\author{A. Gonoskov}
\affiliation{Department of Physics, Chalmers University of Technology, G\"{o}teborg, Sweden}
\affiliation{Institute of Applied Physics, Russian Academy of Sciences, Nizhny Novgorod 603950, Russia}
\affiliation{Lobachevsky State University of Nizhni Novgorod, Nizhny Novgorod 603950, Russia}
\author{M. Marklund}
\affiliation{Department of Physics, Chalmers University of Technology, G\"{o}teborg, Sweden}
\begin{abstract}
Advances in the generation of relativistic intensity pulses with wavelengths in the X-ray regime, through high harmonic generation from near-critical plasmas, opens up the possibility of X-ray driven wakefield acceleration. The similarity scaling laws for laser plasma interaction suggest that X-rays can drive wakefields in solid materials providing TeV/cm gradients, resulting in electron and photon beams of extremely short duration. However, the wavelength reduction enhances the quantum parameter $\chi$, hence opening the question of the role of non-scalable physics, e.g., the effects of radiation reaction. Using three dimensional Particle-In-Cell simulations incorporating QED effects, we  show that for the wavelength $\lambda=5\,$nm and relativistic amplitudes $a_0=10$-100, similarity scaling holds to a high degree, combined with  $\chi\sim 1$ operation already at moderate $a_0\sim 50$,  leading to photon emissions with energies comparable to the electron energies. Contrasting to the generation of photons with high energies, the reduced frequency of photon emission at X-ray wavelengths (compared to   at optical wavelengths) leads to a reduction of the amount of energy that is removed from the electron population through radiation reaction. Furthermore, as the emission frequency approaches the laser frequency, the importance of radiation reaction trapping as a depletion mechanism is reduced, compared to at  optical wavelengths for $a_0$ leading to similar $\chi$. 
\end{abstract}
\maketitle

\section{Introduction}

The generation of a wake structure by the interaction of an optical laser pulse with an underdense plasma has  been demonstrated to accelerate electrons to GeV energies.\cite{TAJIMAW,Esarey,MALKA,LEEMANS} For sufficiently high laser amplitudes, the laser pulse expels the electrons from a region, hence forming a cavity that traps electrons which are accelerated to high energies and oscillate in the transverse field structure, emitting photons  in the X-ray and $\gamma$ range. For optical laser wavelengths, e.g., $\lambda=800\,$nm, the critical plasma density $n_c=m_e\omega^2/4\pi e^2=1.7\times 10^{21}\,$cm$^{-3}$, where $m_e$ is the electron mass, $e$ is the electron charge and $\omega$ is the angular frequency of the laser radiation. This density is orders of magnitude lower than that of solid density materials, which renders solids unfeasible for wakefield acceleration using contemporary optical laser systems. Instead, wakefields driven by optical lasers conventionally use gaseous targets of lower density. On the other hand, inferring from scaling relations, for wavelengths in the X-ray regime, solid materials provide an avenue towards a new generation of even more compact wakefield accelerators -- with TeV/cm gradients and ultra short electron and photon beams, with reduced beam size and emittance\cite{XIAOMEI}. As we will show, the short wavelength enhances the quantum parameter $\chi$. This could  lead to a highly efficient radiation source of high energy photons even at moderate relativistic amplitudes $a_0=eE/m_ec\omega$, where $c$ is the speed of light and $E$ is the electromagnetic field strength. 

The challenge of generating  wakefields in solid materials is closely tied to the feasibility of generating coherent X-ray pulses of sufficient amplitude.  The relativistic electron spring model\cite{RES,RES2} (RES)  is one suggestion of a method which may  provide a source of  high harmonics, which is suitable in realizing  an  X-ray driven wakefield accelerator. The RES model is capable of  describing the interaction of  laser light with  moderately overdense plasmas ($1\leq S<10$, where $S=n_e/a_0n_c$ and $n_e$ is the electron density). Because of the moderate density, energy is stored in the plasma fields and then re-emitted: generating an attosecond burst maintaining relativistic intensity\cite{XUV}. This is different from high harmonic generation in the $S\gg 1$ regime, which is  described by the relativistically oscillating mirror\cite{ROM} (ROM) model, characterized by instantaneous energy balance between the laser and plasma fields (at an apparent point of reflection). This results in an amplitude restriction on the reflected pulse, as it  never can exceed  that of the incoming radiation.  Moreover, the amplitude of the X-ray burst, as described by RES, can be further  increased by using a groove-shaped target\cite{RES}, reaching an increased range of relativistic amplitudes  $a_0\sim 10$ for parameters available at current laser facilities.

Similarity theory\cite{SIMI,Bubble} aims at reducing the vast parameter space, comprising laser plasma interaction, into an as small parameter set as possible from which the general properties of the plasma dynamics and acceleration process can be derived.
Consider a plasma with electron density $n_e$ interacting with a pulse propagating in the $x$-direction with vector potential of the form $\g a(x,y,z,t)=a(r_\perp/R,(t-x/c)/\tau)\cos(\omega t-kx)\hat z$, where  $r_\perp$ is the transverse radial coordinate, $R$ is the width of the pulse, $\tau$ is its duration and $k=\omega/c$ is the wave number. The interaction can be fully described by the four dimensionless parameters $kR$,  $\omega\tau$, $n_e/n_c$ and $a_0$. 
For high $a_0$, the electrons become relativistic, moving at a speed close to the speed of light and the four dimensionless parameters can be reduced by combining  $n_e/n_c$ and $a_0$ into the similarity parameter $S=n_e/a_0n_c$, resulting in the three similarity parameters\cite{SIMI} $S$, $\hat R={S}^{1/2} kR$ and $\hat \tau={S}^{1/2}\omega \tau$. This description in terms of similarity parameters suggest that a wakefield regime in principle can be realized using short wavelength pulses (X-rays) and high density (solid) targets, similar to what currently is realized with optical wavelengths and lower density targets. For an X-ray pulse with $\lambda=5\,$nm, the critical density is $n_c=4.5\times 10^{25}\,$cm$^{-3}$ rendering $S<1$ at moderate $a_0$ for solid density materials (charcterized by electron densities $5\times 10^{22}\,$cm$^{-3}<n_e<5\times10^{24}\,$cm$^{-3}$). From the similarity scaling relations it follows that for  a fixed $S$ parameter, the current density in the bunch of trapped electrons scales as $1/\lambda^2$, whereas the current is unchanged by rescaling the wavelength, and the trapped charge is proportional to $\lambda$. The enhanced current density   reflects  the increase of the critical density, followed by that the  current is unchanged  as the increase of the current density is compensated by the reduction of the transverse scales and finally   the trapped charge is reduced due to the  reduction of the dimension of the wake along the propagation direction. On the other hand, the electron energies only scale with $a_0$ and are consequently independent of the wavelength.

\begin{figure*}[!htb]
\centerline{\includegraphics[width=1.2\textwidth]{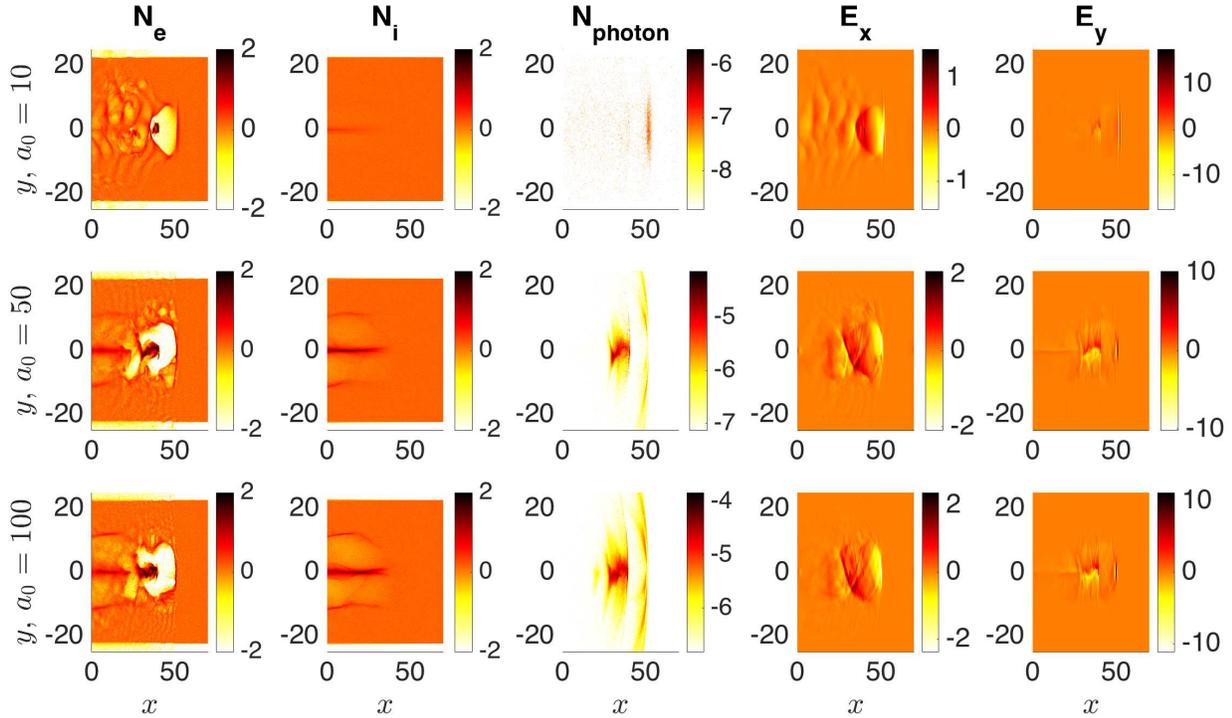}}
\caption{\label{fig::1}Electron, ion and photon densities as well as longitudinal and transverse electric fields at time $ t=100T$,  for $a_0=10$ (upper row), 50 (middle row) and 100 (bottom row).   The densities are in logarithmic scale in units of the background $n_e=Sn_ca_0$ and the spatial scale is in units of $\lambda$. Finally, the longitudinal and transverse electric fields are given in units of $m_ecS^{1/2}a_0\omega/e$. }
\end{figure*}

Early work utilizing  X-rays to drive slow-wave accelerating structures in  metallic crystal channels was conducted by Tajima {\it et al.}\cite{TAJIMAXC} in the 1980s.  More recently the concept of X-ray driven wakefield acceleration in solids has been revived by Zhang {\it et al.}\cite{XIAOMEI}, with 2D Particle-In-Cell simulations of wakefields driven by X-rays of  $\lambda=1\,$nm and moderate relativistic amplitudes $a_0\sim 1$-$10$.  In this regime, they concluded good scalability of the accelerating structures, but with more prominent radiation reaction due to the  enhancement of the quantum parameter $\chi=\gamma E_\perp/E_c$, which governs the  probability distribution and rate for photon emission. Here, $\gamma$ is the gamma-factor of an electron, $E_\perp$ is the transverse electric field and $E_c=m_e^2c^3/e\hbar$ is the Schwinger field.  As $\chi\approx  a_0^2(\lambda_c/\lambda)$ (with $\lambda_c=2.4\times 10^{-3}\,$nm being the Compton wavelength) and that $\chi$ in the semi-classical limit ($\chi\ll 1$) represents the quotient between the typical photon energy and electron energy, X-ray driven wakefields show promise as a source of high energy photons. The scaling of $\chi$ indicates that the quantum regime for radiation reaction (characterized by $\chi\sim 1$) can be probed already at moderate $a_0<100$.

Motivated by the prospects of probing $\chi\sim 1$, we extend the analysis of X-ray  wakefield acceleration to moderate relativistic intensities $a_0\sim 10 -100$. Relativistic X-rays obtained by  harmonic generation from laser--solid interactions have been chosen as a motivation for this work. However, this is not a restriction of the results in the paper, which merely require a coherent X-ray pulse of sufficient amplitude, independent of source. We present results from 3D  simulations of X-ray driven wakefield acceleration, using the  Particle-In-Cell (PIC) code ELMIS\cite{elmis}, implementing QED effects\cite{QED}. We adress  the scalability of the accelerating structure, and effect on electron acceleration from operating in the strong field QED regime already at moderate relativistic amplitudes, hence deducing the prospects  and limitations  of this regime. 

\section{Method and setup}

Simulations are performed using the three dimensional PIC-code  ELMIS\cite{elmis} (Extreme Laser-Matter Interaction Simulator), which is based on a spectral method for solution of Maxwells equations. The method introduces no dispersion error, neither due to  grid or time stepping, which is important for accurate modelling of wakefield acceleration. Furthermore, ELMIS implements an adaptive event generator\cite{QED} to efficiently handle the disparity of scales introduced by QED-effects. A coherent linearly polarized Gaussian X-ray pulse with wavelength $\lambda=5\,$nm interacts with a plasma of  charge to mass ratio $Z/A=0.5$ and density such that $S=10^{-3}$. For the studied range of relativistic  amplitudes: $a_0=10$-100, this choice of $S$ parameter gives densities within the solid range. 
The corresponding laser intensities are given by $I=1.4\times 10^{24}\times a_0^2/\lambda_{\text{nm}}^2\,$W/cm$^2$, and are in the range  $5.5\times 10^{24}\,$W/cm$^2$  to $5.5\times 10^{26}\,$W/cm$^2$. 
 The duration (FWHM) of $3T$, where $T$ is the laser period, is determined by  output from PIC-simulations of the RES-mechanism\cite{XUV}. A Gaussian pulse is chosen in order to maintain a generic setup and can be obtained by filtering the RES-pulse, which has a sawtooth-shape. Filtering can be motivated by that the pondermotive force of the pulse is connected to its envelope. The width of the laser pulse corresponds to the matching criterion $k_pR \sim \sqrt{a_0}$ (equivalent to $\hat R=1$). The plasma density is first linearly increased to $2n_e$ (where $n_e=Sa_0n_c$) over a distance of $2\hat R$ and then linearly decreased to $n_e$ over a distance of $2\hat R$, hence allowing for density gradient injection. 

The use of X-ray wavelengths makes it necessary to validate the use of a collisionless model.  For relativistic electrons, the (electron-ion) collision frequency is given by $v_{ei}=4\pi n_e Zc r_0^2\log\Lambda/\gamma^2$, where $Z$ is the ion charge number, $\log\Lambda$ is the Coulomb logarithm and $r_0$ is the classical electron radius. Using normalized units: $\gamma= \hat \gamma a_0$, the collision frequency can be expressed as $v_{ei}=4\pi n_c SZc r_0^2\log\Lambda/(\hat \gamma^2a_0)\sim 1/\lambda^2$. Taking into account that the time scale for the acceleration process is normalized with respect to the laser period and hence scales as $\lambda$, the wavelength dependence of the importance of collisional processes is found to scale as $\sim1/\lambda$. For the studied parameter range and values $Z\sim 10$, $\log\Lambda\sim 10$, the collision frequency becomes $v_{ei}=4\pi n_c SZc r_0^2\log\Lambda/\hat \gamma^2a_0=2.2\times 10^{-6}/(T\hat \gamma^2a_0)\ll 1/T$, for $\hat \gamma\sim 1$. Hence, collisional processes can to first order be neglected.

The scalability of the laser-plasma interaction also relies on that the wavelength is long enough to not probe the configuration of single particles. Specifically, the dimensions of the ion lattice structure identifies a lower wavelength limit for the validity of scaling the dynamics from optical wavelengths. Assuming parameters $Z\sim 10$, $n_e\sim 5 \times 10^{24}$\,cm$^{-3}$, the average separation of ions can be estimated by $d=(n_e/Z)^{-1/3}\sim 0.1\,$nm.   Hence, the wavelength $\lambda=5\,$nm remains well above the ion separation distance, motivating that effects due to lattice structure can be neglected. Furthermore, a comparison of the field strength from individual ions and the strength of the transverse electric field in the wake shows that the latter is dominant. 

The simulations are performed with a moving box of size  $l_x=70\lambda$ discretized using 512 cells in the laser propagation direction and $l_\perp=10\hat R=50.3\lambda$ using 256 cells in the transverse direction.   Simulations are performed with two particles per cell.

\section{Results}

\subsection{Wake structure}

Figure \ref{fig::1} shows electron, ion and photon densities as well as transverse and  longitudinal electric fields  at time $t=100T$  for the cases with $a_0=10$, 50 and 100. At these intensities, the laser expels the electrons leading to the formation of a bubble. The inclusion of QED-effects, such as radiation reaction, is found to have a negligible effect on the wake. We conclude that the wake structure is scalable from optical wavelengths, with a reduction of the dimensions of the accelerating structure proportional to $\lambda$ and an enhancement of the field strengths proportional to $1/\lambda$. 

\begin{figure}[!htb]
\includegraphics[width=.48\textwidth]{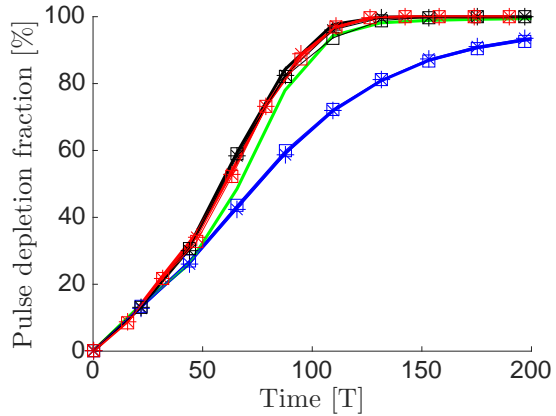}
\caption{\label{fig::2}Pulse depletion as a function of time for  $a_0=10$ (blue), 20 (green), 50 (red) and 100 (black). The cases $a_0=10$, 50 and 100 are also shown with radiation reaction disabled ($*$-marker), and with radiation reaction at wavelength $\lambda=800\,$nm (square marker). Both in the X-ray driven and optical cases: radiation reaction has a negligible effect on pulse depletion for the considered range of intensities.}
\end{figure}

Although the structure of the wake is scalable from optical wavelengths, the onset of ion motion for  $a_0$ approaching one hundred breaks the $S$-number similarity. The reduction of $n_e/n_c$ and $a_0$ into the similarity parameter $S$  relies on  that the velocity $\vec v$ can be approximated by $\vec v/c\approx\vec p/|\vec p|$,
which holds  if  $(m/m_ea_0)^2<1$, for all mobile species of mass $m$. This is the case for electrons irradiated by moderate relativistic amplitudes, but  rarely for  ions.  Instead, we operate in  the regime $1\ll a_0\ll m_i/m_e$, for which $S$-number similarity does not hold and  $a_0$  must be specified to describe the dynamics. Figure \ref{fig::2} shows the time dependence of the fraction of the pulse energy which has been transferred to other energy channels, for different $a_0$, both with and without radiation reaction. The onset of ion motion leads to an increased depletion rate, whereas radiation reaction (in this regime) is confirmed have a negligible effect on depletion. The figure also shows result for optical wavelengths, which are in agreement with those from the X-ray driven regime.

The character of the ion motion is similar to that discovered by Wallin {\it et al}\cite{WALLIN}: the ion density in the bubble is reduced by radiation pressure expelling ions, leading to an increased ion density at the bubble contour. The reduction is enhanced by that the ions inside the bubble strongly interact with the trapped electron bunch, thereof forming an ion column along the path traced by the electron bunch. As a consequence, the wake structure does not close as would be expected in the bubble regime when ion motion is disregarded. Ion motion leads to a less stable wake structure, with more complicated transient field structure, resulting in  less efficient focusing of the electron bunch compared to the dynamics with immobile ions. As is witnessed in Figure \ref{fig::1}, this  also affects the angular spread of the emitted photons.

\begin{figure}[!htb]
\includegraphics[width=0.5\textwidth]{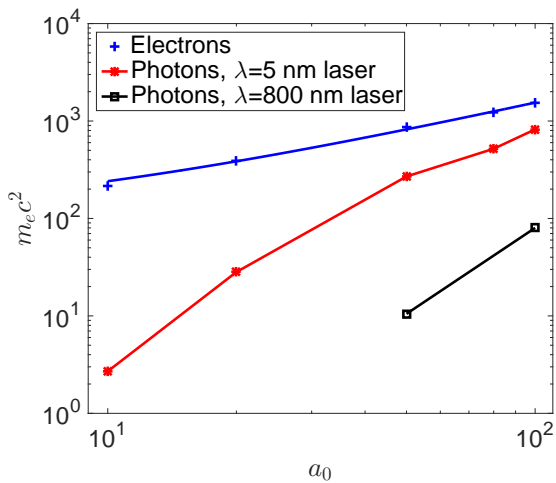}
\caption{\label{fig::3}Maximum energy (at time $t=100T$) for electrons and photons as a function of $a_0$. The electron energies are fitted to a linear curve which show good agreement with the data,  as predicted by similarity theory. For $a_0\sim 100$, the maximum photon energy is approximately half of the maximum electron energy, indicating that the effect of recoils may be significant. The lower photon energies obtained at the optical wavelength $\lambda=800\,$nm are indicated for comparison.}
\end{figure}

\begin{figure*}[!htb]
\includegraphics[width=1\textwidth]{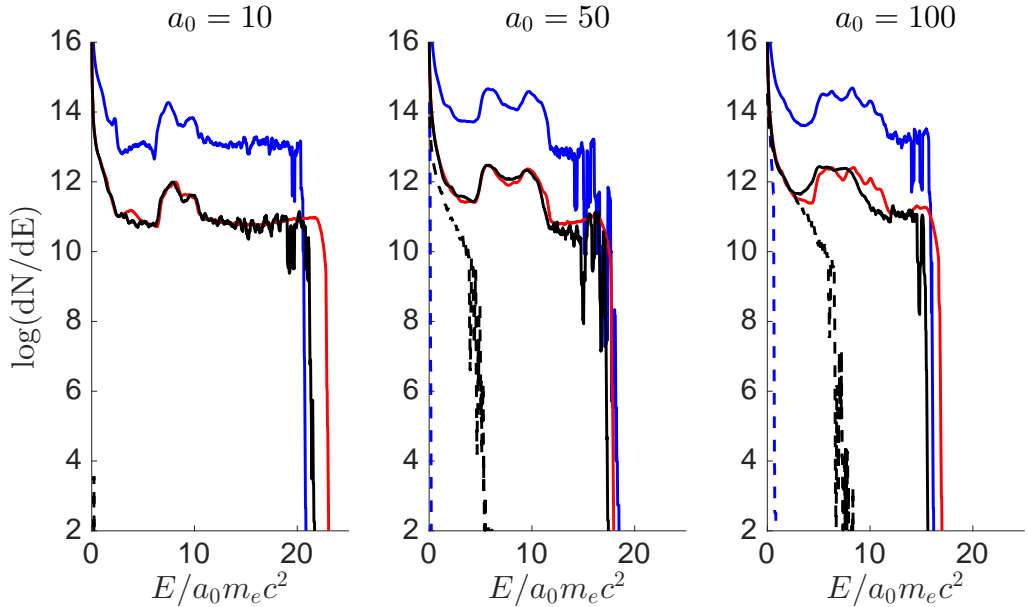}
\caption{\label{fig::4}Energy spectrum for electrons (lines) and photons (dashed lines) at time $ t=100T$, for the relativistic amplitudes $a_0=$10, 50 and 100 in X-ray driven cases with (black) and without (red) radiation reaction, as well as cases with an optical wavelength driver ($\lambda=800\,$nm, blue).}
\end{figure*}

\subsection{Scaling of electron beam and radiation generation}

Figure \ref{fig::3} shows the maximum energy of electrons and photons at time $ t=100T$. Electron energies scale linearly with $a_0$, as predicted by similarity scaling laws. The maximum photon energy shows a strong dependence on $a_0$: increasing from a negligible fraction of the electron energy for $a_0=10$ up to comparable levels for $a_0$ exceeding 50. In the classical regime, $\chi$ can be interpreted as the quotient of the typical photon energy to the electron energy:  $2\hbar\omega/3m_ec^2\gamma$. On the other hand, in this regime, where $\delta=\hbar\omega/m_ec^2\gamma$ is close to unity, quantum corrections to the radiation reaction are essential and $\chi$ is in general not proportional to $\delta$. Therefore, we choose to estimate $\chi$  from the  electron energies and transverse electric fields. In terms of normalized coordinates ($\hat E_\perp$ and $\hat \gamma$): $\chi=1.5\times 10^{-5} \hat \gamma \hat E_\perp  a_0^2$. Estimating $\hat \gamma\sim 15$ and $ E_\perp\sim 5$  from Figures \ref{fig::1} and \ref{fig::4}: $\chi\approx 1.2a_0^2\times 10^{-3}  =0.1$, 0.4, 2.8, 7.0 and 11.0 for $a_0=10$, 20, 50, 80 and 100. In summary, the X-ray wakefield accelerator is not only a compact source of GeV electrons (for $a_0=100$), but also has high values of $\chi$ yielding photons of similar energies.

Figure \ref{fig::4} shows  energy spectra for electrons and photons at time $ t=100T$, for the three cases with $a_0=10$, 50 and 100,  with and without radiation reaction, as well as for the optical wavelength $\lambda=800\,$nm. Radiation reaction is found to have a minor effect on the energy spectrum. From similarity scaling,  the energy spectra are predicted to scale as ${\text{d}N}/{\text{d}\epsilon}=Sn_c\lambda^3{\text{d}\hat N}/{\text{d}\hat\epsilon}$, where ${\text{d}\hat N}/{\text{d}\hat\epsilon}$ is the energy spectra in similarity normalized coordinates. Since $n_c\sim 1/\lambda^2$, the   wavelength dependence becomes ${\text{d}N}/{\text{d}\epsilon}\sim \lambda$. Notice that the pulse energy also  scales as $\sim\lambda$. The reduction of the amount of available energy is hence manifest in a reduction of the accelerated particle number, but not in the resulting energies. This is verified by comparing the spectra at X-ray and optical wavelength and is different from the effect of increasing $a_0$, where the quadratic increase in pulse energy is distributed into a linear increase of both the accelerated particle number and the achieved energies. 

\begin{figure*}[!htb]
\includegraphics[width=\textwidth]{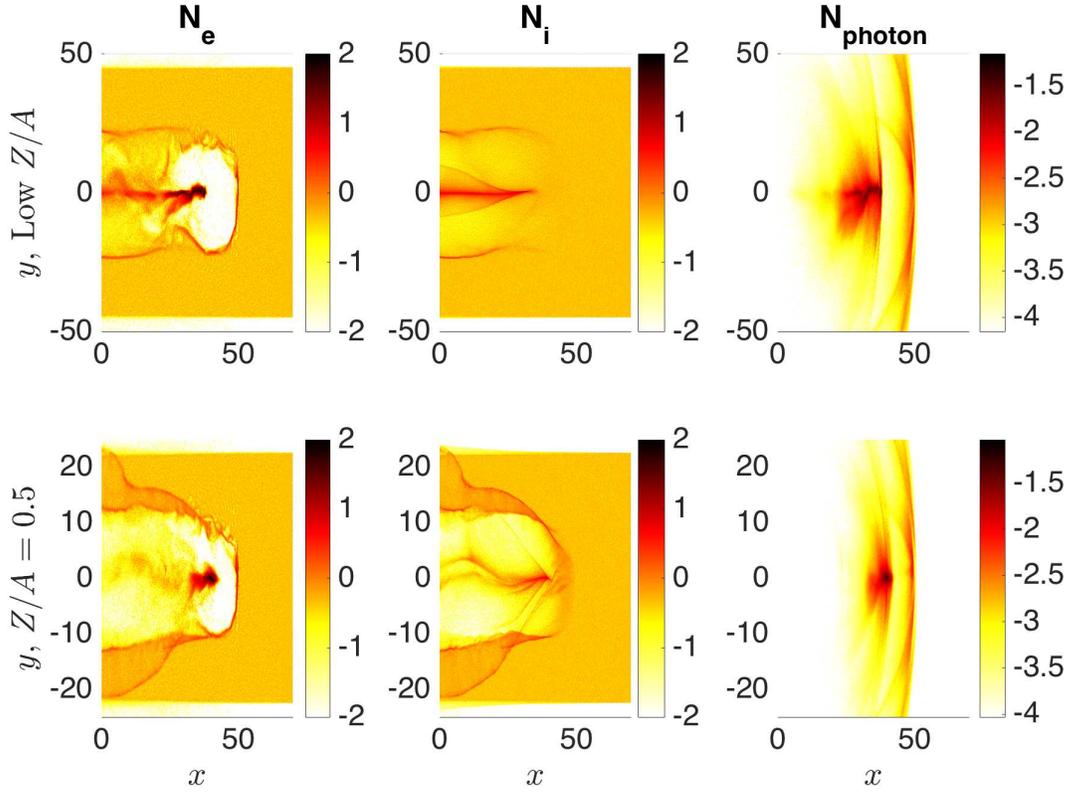}
\caption{\label{fig::5}  Electron, ion and photon densities at time $t=100T$ for the cases with $a_0=1265$ and $Z/A=1/25.3$ (first row) as well as $a_0=1265$ and $Z/A=0.5$ (second row). The densities are in logarithmic scale in units of the background density $n_e=Sn_ca_0$ and the spatial scale is in units of   $\lambda$.}
\end{figure*}

\begin{figure}[!htb]
\includegraphics[width=0.5\textwidth]{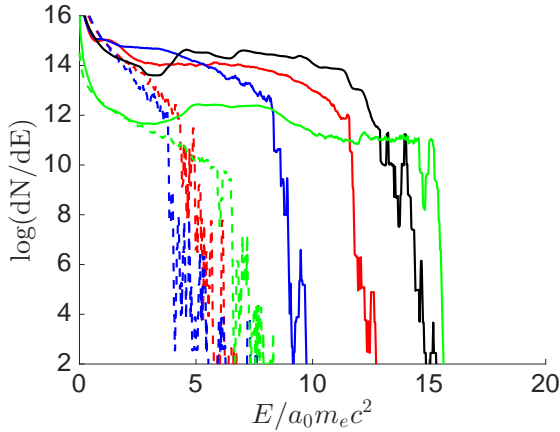}
\caption{\label{fig::6}Electron (lines) and photon (dashed lines) spectra at time $t=100T$ for the X-ray driven case with $a_0=100$ (green), as well as  optical cases ($\lambda=800\,$nm) with (1) $a_0=1265$, $Z/A=1/25.3$, with QED effects  (blue), (2) $a_0=1265$,  $Z/A=1/25.3$, without QED effects (black) and (3) $a_0=1265$, $Z/A=1/2$, with QED effects  (red). }
\end{figure}

\begin{figure}[!htb]
\includegraphics[width=0.5\textwidth]{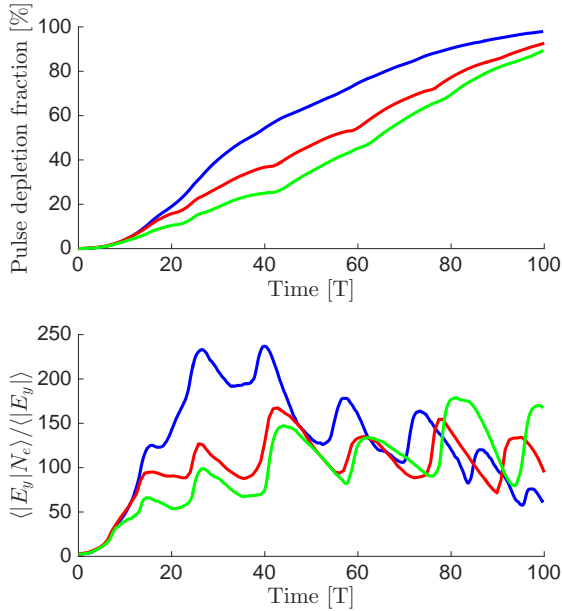}
\caption{\label{fig::7}Upper: Fraction of the pulse  which has been depleted as a function of time for the X-ray driven case with $a_0=316$ (red), as well as the optical case ($\lambda=800\,$nm) with  $a_0=4000$ and $Z/A=1/25.3$, with (blue) and without (green) QED-effects.  Lower: Measure of the overlap of the laser pulse and electron density as a function of time. }
\end{figure}

To assess the effect of the wavelength on the spectral properties for photon emission, consider the emission probability  \cite{QED}    
\begin{gather*}
\frac{\text{d}P}{\text{d}\delta}=\\\left[\Delta t\frac{e^2m_ec}{\hbar^2}\right]\frac{\sqrt{3}}{2\pi}\frac{\chi}{\gamma}\frac{1-\delta}{\delta}\left\{F_1(z_q)+\frac{3}{2}\delta\chi z_q F_2(z_q)\right\}
\end{gather*}
where $\delta$ is the photon to electron energy quotient defined above, $z_q={2\delta}/3\chi (1-\delta)$ and $F_1(x)$, $F_2(x)$ are the first and second Synchrotron functions. Expressing the pre-factor in terms of nomralized coordinates:  $\chi=\hat\chi \, a_0^2({\lambda_\text{c}}/{\lambda})$, $\gamma=a_0\hat\gamma$ and $\text{d} t=T\text{d}\hat t$, the probability distribution takes the form
\begin{gather*}
\frac{\text{d}P}{\text{d}\delta}=\\\sqrt{3}a_0\,\Delta \hat t\,\frac{e^2}{c\hbar}\frac{\hat\chi }{\hat\gamma}\frac{1-\delta}{\delta}\left\{F_1(z_q)+\frac{3}{2}\delta\chi z_q F_2(z_q)\right\}.
\end{gather*}
For $a_0$ chosen to preserve  $\chi$, the spectral shape is the same although the emission frequency scales as $\sim \sqrt{\lambda}$. Using an X-ray pulse to operate a wakefield at a prescribed $\chi$ will hence result in fewer emitted photons per electron than at optical wavelengths. The reduction of the number of emitted photons in the X-ray driven case is further amplified by the decrease of the number of accelerated electrons: both because of the scaling of the electron spectrum with the wavelength and  $a_0$ (which is increased in the optical case to achieve the same $\chi$).

The  total radiated intensity of an electron is  $I_r=0.67  I_0\chi^{2}$ in the classical regime and $0.37 I_0\chi^{2/3}$ in the quantum regime, where $I_0={e^2m_e^2c^3}/{\hbar^2}$. Hence, integrating $I_r$ along the trajectories of all particles, leads to a total radiated energy that scales as $a_0n_c\lambda^4\bar I_r\sim a_0\lambda^2 \bar I_r$, where $\bar I_r$ is an average value only depending on $a_0^2({\lambda_c}/{\lambda})$. Comparing to the total energy in the pulse which scales as $\lambda a_0^2$, the fraction of energy which goes into photon emission equates to $\lambda \bar I_r/a_0$. If $a_0$ is chosen to match  $\chi$ at X-ray and optical wavelengths, the energy fraction which is lost through radiation reaction  is predicted to scale as $\sqrt{\lambda}$, which  is consistent with the scaling of the photon emission probability. 

The non-perturbative approach in the previous paragraphs to the impact of radiation reaction  is  only valid  for $\chi\ll 1$ and consequently small $\delta$. These conditions hold for $a_0\sim 10$ but not $a_0\sim 100$, where comparable photon and electron energies indicate that  single recoil events can significantly alter the path of electrons. Radiation reaction therefore has not only  a significant effect on the dynamics of the trapped electrons, but this effect is also different for X-ray compared to optical wavelengths. To illustrate some differences, we consider a setup with wavelength $\lambda=800\,$nm and $a_0=1265$. In a perturbative setting, this leads to the same $\chi$ as in the X-ray case with $a_0=100$.  We consider the following  cases: one with $Z/A=0.5$ and one where the ion mass has been increased by a factor 12.65, i.e.  $Z/A=1/25.3$.  In the latter case, we perform simulations both with and without QED-effects. Figures \ref{fig::5} and  \ref{fig::6} show snapshots of electron, ion and photon densities, as well as  energy spectra  at time $t=100T$.  Comparing to the X-ray driven case with $a_0=100$, increased radiation reaction leads to decreased electron energies, with respect to similarity normalized coordinates, in the  optical case. Furthermore, the ion dynamics at $Z/A=0.5$ is widely different compared to if the smaller charge to mass ratio $Z/A=1/25.3$ is used, which corresponds to operating with the same ion relativistic amplitude as for the X-ray driven  case. In summary, the X-ray regime allows for generation of high energy photons with reduced energy losses from the electron population due to radiation reaction and smaller influence of ion motion than in the optical regime, when choosing $a_0$ to match $\chi$.

Finally, the lower $a_0$ and more stochastic nature of photon emission in the X-ray regime may reduce the effect of radiation reaction trapping, i.e. the trapping of electrons due to radiation friction in the most intense part of the laser field\cite{RRT}. These electrons then radiate,  creating a channel for energy loss and faster depletion of the laser pulse. 
For large $a_0$, the expected time between photon emissions can be estimated by $\tau\approx 31 T\left({\lambda_c}/{\lambda a_0} \right)^{1/3}$, yielding $\tau\approx 2.5 T/a_0^{1/3}$ for $\lambda=5\,$nm.  Hence, the time between photon emissions in the X-ray regime decreases from $1.2T$ to $0.5T$ when  $a_0$ increases from 10 to 100, which is comparable to the laser period and hence expected to be unsuitable for trapping particles. The situation is different for optical wavelengths, where $\tau\approx 0.46 T/a_0^{1/3}$ (for $\lambda=800\,$nm), which gives $0.2T$ to $0.1T$ for the same range of relativistic amplitudes. Adjusting for operation at the same $\chi$  leads to a  further reduction of the emission time by a factor 2.3. Figure \ref{fig::7} shows
the time dependence for depletion of the  pulse in the X-ray driven case with $a_0=316$, $Z/A=0.5$, and optical case with $a_0=4000$, $Z/A=1/25.3$, which corresponds to the same $\chi$. Simulations results are shown both with and without QED-effects. Faster depletion is observed for the optical case with QED-effects than in the X-ray driven case, where the depletion rate more closely follows that without QED-effects. The figure  also shows the overlap of the electron density and the laser pulse as a function of time. The overlap is found to be larger in the optical case if QED-effects are included, which indicates that radiation reaction trapping occurs in the optical case. On the other hand, the overlap in the X-ray case is similar to the case without QED-effects and gives  hence no indication of radiation reaction trapping.

\section{Conclusions}

The indication that it is possible to  generate relativistic intensity high harmonics has opened up the prospects for wakefield acceleration in solid materials. Similarity scaling laws, summarized in Table \ref{TAB1}, are  found to hold to good approximation, creating  ultrashort electron and photon beams, with spatial and temporal scales reduced proportionally to the wavelength reduction. The reduced transverse scales  are  compensated by the enhanced density, leading to a preserved current although the total number of accelerated particles  is reduced proportionally to the wavelength. The X-ray driven wakefield is therefore an excellent method for applications demanding ultrashort electron and photon beams. 

X-ray driven wakefields also show potential as sources of high energy photons,  with operation in the $\chi\sim 1$ regime already at moderate $a_0\sim 50$, leading to electrons and photons of similar energy. At the same time, the reduced frequency of photon emission (in relation to the source period) leads to  smaller losses  of energy from the electron population (than at optical wavelengths with similar $\chi$) through radiation reaction. Combining these aspects, radiation reaction has a perturbative effect associated with large but infrequent recoils for the accelerated electron population.

When comparing the high $\chi$ regime for wakefields driven by X-rays and at optical wavelengths, a major difference arises due to ion motion. Although ion motion is scalable between X-ray and optical wavelengths, transforming $a_0$ to preserve $\chi$ significantly changes the ion dynamics.  For  moderate $a_0$, as considered in this paper, the ion motion in the X-ray regime is modest. On the other hand, when transformed to preserve $\chi$ to the optical regime, the radiation pressure also pushes  the ions to relativistic speeds, creating a widely different regime of interaction.  

In addition to radiation reaction, pair production will in the limit of high $a_0$ play a role and present a potential roadblock for extreme wakefield acceleration. The formation of cascades in the wake followed by scattering of the positrons from the electron potential well or cascades  in the laser field may have a destructive effect on the accelerating structure. In the currently studied range of $a_0$ only trace amounts of pairs were observed. However, the quantum number for  pair-production is expected to be enhanced in the X-ray driven regime, with the  scaling   $\eta\sim 1/\lambda^2$ for  $\chi\ll 1$ and  $1/\lambda$ in the high $\chi$ regime. Details of this remain a topic for future studies. 

The challenges of generating coherent high amplitude high harmonics makes the $a_0\sim 1$ regime the first step towards realizing an X-ray wakefield accelerator. In this regime, ions are relatively immobile, as is evident from the case with $a_0=10$. For  wavelengths $\sim5\,$nm, the average distance between ions remains  comparatively small, motivating the approximation of ions with an average density. For  shorter wavelengths, the ion lattice structure may influence and potentially guide the wake structure, but is left for future studies.

\begin{table}[tb]
\caption{\label{TAB1}Scaling relations of wakefield acceleration for $\lambda$ and $a_0$ while keeping the $S$ number  fixed. The upper part of the table contains scalings from similarity theory, whereas the lower part contains scalings for  QED effects. Here, $\alpha=2$ or $\alpha=2/3$, depending on if the electron motion is characterized by that $\chi$ is smaller or bigger than unity. }
\begin{tabular}{l rl rl  }
\hline\hline
 Quantity & Scaling \\ \hline
Timescale& $\lambda$& \\
Spatial dimensions of cavity& $\lambda$& \\
Density & $a_0/\lambda^2$\\
Current density & $a_0/\lambda^2$\\
Current & $a_0$\\
Trapped charge & $\lambda a_0$\\
Electron energy & $a_0$\\
Electromagnetic fields   & $a_0/\lambda$    \\
Pulse energy & $\lambda a_0^2$\\
Ion motion & $a_0$    \\
\hline
Quantum parameter, $\chi$          &    $a_0^2/\lambda$   \\ 
Radiated energy fraction& $\lambda^{1-\alpha}a_0^{2\alpha-1}$\\
Photon energy  ($\chi\ll1$)&$a_0^3/\lambda$ \\
Photon energy  ($\chi \sim 1$)&$a_0$\\
Time between emissions ($\chi \ll 1$) & $T/a_0$\\
Time between emissions ($\chi \sim 1$) & $T/(\lambda a_0)^{1/3}$
 \\ \hline\hline
\end{tabular}
\end{table}

\section{Acknowledgements}

This research was supported by the Knut \& Alice Wallenberg Foundation Grant Plasma based compact ion sources, and the Swedish Research Council (grant 2016-03329). The simulations were performed on resources provided by the Swedish National Infrastructure for Computing (SNIC) at High Performance Computing Center North (HPC2N). 

\bibliography{XWFA}

\end{document}